\documentclass[a4paper, 12pt]{article}
\usepackage{subcaption}
\usepackage{tikz}
\usetikzlibrary{arrows,decorations.markings,cd}
\usepackage[normalem]{ulem}
\usepackage{latexsym,amsmath,amsfonts,amssymb}
\usepackage[latin1]{inputenc}
\usepackage[american]{babel}
\usepackage{bbm}
\usepackage[nosort]{cite}
\usepackage{hyperref}

\begin{document}

\thispagestyle{empty}
\begin{flushright}
\end{flushright}

\begin{center}
\vskip 10mm
{\huge  Notes on $\,\mathcal{N}=1$ QCD$_3$ \\[.5em]with baryon superpotential
}
\\[15mm]
{Vladimir Bashmakov$^{1}$, Hrachya Khachatryan$^{2}$}
 
\bigskip
{\it
$^1$ Dipartimento di Fisica, Universit\`a di Milano-Bicocca and INFN, Sezione di Milano-Bicocca, I-20126 Milano, Italy\\  
$^2$  International School of Advanced Studies (SISSA), Via Bonomea 265, 34136 Trieste, Italy\\[.5em]
}

{\tt vladimir.bashmakov@unimib.it,  hrachya.khachatryan@yahoo.com}

\bigskip
\bigskip

{\bf Abstract}\\[5mm]
{\parbox{14cm}{\hspace{5mm}

We study the infrared phases of $\mathcal{N}\,=\,1$ QCD$_3$ with gauge group $SU(N_c)$ and with the superpotential containing the mass term and the baryon operator. We restrict ourselves to the cases, when the number of flavors is equal to the number of colours, such that baryon operator is neutral under the global $SU(N_F)$. We also focus on the cases with two, three and four colours, such that the theory is perturbatively renormalizable. On the $SU(2)$ phase diagram we find two distinct CFTs with $\mathcal{N}\,=\,2$ supersymmetry, already known from the phase diagram of $SU(2)$ theory with one flavor. As for the $SU(3)$ phase diagram, consistency arguments lead us to conjecture that also there supersymmetry enhancement to $\mathcal{N}\,=\,2$ must occur.  Finally, similar arguments lead us to conclude that also for the $SU(4)$ case we should observe supersymmetry enhancement at the fixed point. Yet, we find that this conclusion would be in contradiction with the renormalization group flow analysis. We relate this inconsistency between the two kinds of arguments to the fact that the theory is not UV complete.
}}
\end{center}

\newpage
\pagenumbering{arabic}
\setcounter{page}{1}
\setcounter{footnote}{0}
\renewcommand{\thefootnote}{\arabic{footnote}}

\section{Introduction}

One of the cornerstones in understanding of any quantum field theory (QFT) is the question about its infrared (IR) phase. If a theory is gapped, and it does not have any observables surviving below certain energy scale, it is said to be in a \textit{trivial gapped phase}. The other possibility is that there is the mass gap, but non-local observables still possess non-trivial dynamics: in this paper we will call this situation a \textit{topological phase}, since this dynamics is going to be described by a topological field theory (TFT). It can also be that some continuous global symmetries are spontaneously broken, leading to the absence of the mass gap and appearance of Goldstone particles in the IR. Of course, there may also be mixed cases, where both Goldstone particles and a TFT coexist. Quite often a theory may possess several phases depending on values of the UV parameters (coupling constants). Then, while smoothly varying the parameters, we can move from one phase to another. If the phase transition is of second order, two phases are separated by a conformal field theory (CFT).

Notable progress in revealing phase diagrams of three dimensional non-supersymmetric gauge theories has been seen in the last few years \cite{Komargodski:2017keh, Gaiotto:2017tne, Gomis:2017ixy, Cordova:2017vab, Cordova:2017kue, Argurio:2018uup, Aitken:2018cvh, Choi:2018tuh, Cordova:2018qvg, Argurio:2019tvw, Armoni:2019lgb, Baumgartner:2019frr, Aitken:2019mtq, Aitken:2019shs, Choi:2019eyl}.
In particular, phases non-visible semiclassically, and whose dynamics is quantum by nature, were conjectured to be present in several models: they were dubbed \textit{quantum phases} in the literature. The existence of such phases followed from considerations of symmetries and related 't Hooft anomalies, including anomalies in discrete and 1-form symmetries (see \cite{Gaiotto:2014kfa, Aharony:2016jvv, Benini:2017dus} and references given above), while their conjectured IR dynamics came mostly as a guesswork. Those conjectures were subject to stringent checks, including the above mentioned 't Hooft anomalies matching, realisation of 3d theories on domain walls of 4d theories and D-brane constructions.

Despite this significant progress, gauge theories without supersymmetry are hard to study, and often one has to refer to an educated guesswork rather than systematic methods. In this regard, thrilling possibilities open up in the realm of $\mathcal{N}\,=\,1$ theories\footnote{In three dimensions $\mathcal{N}\,=\,1$ supersymmetry is generated by two real supercharges, combining into a single Majorana spinor.}. These theories  recently have attracted much attention in the literature \cite{Bashmakov:2018wts, Benini:2018umh, Gaiotto:2018yjh, Eckhard:2018raj, Benini:2018bhk, Choi:2018ohn, Fazzi:2018rkr, Bashmakov:2018ghn, Rocek:2019eve, Bandos:2019qok, Aharony:2019mbc, Ghim:2019rol}. They are not as constrained as their cousins with higher supersymmetry, in particular $\mathcal{N}\,=\,1$ supersymmetry is "real", thus does not give rise to non-renormalization theorems, relying on holomorphy. This lack of protection makes quantum effects ubiquitous and important for the dynamics. Still, presence of (even little amount of) supersymmetry provides certain useful tools that allow for better control and deeper understanding of these theories.

A tool that has been playing a major role so far is the renowned Witten index \cite{Witten:1982df}. It counts the difference between the number of bosonic ground states and the number of fermionic ground states, when a theory is quantized on a torus, and can be defined for any amount of supersymmetry. The Witten index does not change upon the smooth variation of the parameters, and can jump only when the superpotential changes its asymptotic behaviour.

Another property universal for all the supersymmetric theories is that all the SUSY vacua have vanishing energy, and as such phase transitions between SUSY vacua are always of second order. These two basic observations served as a guiding principle exploited in \cite{Bashmakov:2018wts}, where phases of $\mathcal{N}\,=\,1$ SQCD with an adjoint matter multiplet and SQCD with a fundamental matter multiplet were described as functions of the matter multiplet mass. The subsequent generalisation for SQCD with several flavors was given in \cite {Choi:2018ohn}.

In this paper we continue the investigation of $\mathcal{N}\,=\,1$ SQCD with gauge group $SU(N_c)$ and with $N_F$ fundamental flavors, and study IR phases of the theory deformed by two superpotential operators: the mass operator considered previously and the baryon operator. We restrict ourselves to the case $N_c\,=\,N_F$, such that baryon deformation breaks explicitly $U(1)_B$, but is invariant under global $SU(N_F)$. The superpotential we consider thus takes the form
\begin{equation}\label{GeneralSuperpotential}
\mathcal{W}\,=\,m\,\text{Tr}\,Q\,\bar{Q}\,+\,(\lambda\,\text{det}Q\,+\,\text{c.c.}).    
\end{equation}
We thereby consider three cases in order.

We start with $SU(2)$ theory with two flavors. In this case baryon operator is quadratic, and so in effect we have two different mass terms\footnote{In this respect, our setup is similar to \cite{Argurio:2019tvw,Baumgartner:2019frr}, where non-supersymmetric QCD with two independent mass terms was discussed.}. We find that two-dimensional phase diagram is crossed by two walls where the Witten index jumps, and by two lines where second order phase transitions occur. The first CFT turns out to be the same as for $SU(2)_{k\,-\,\tfrac{1}{2}}$ with one flavor, and the second CFT is the same as for $SU(2)_{k\,+\,\tfrac{1}{2}}$, again with one flavor. It is known \cite{Choi:2018ohn, Bashmakov:2018ghn, Avdeev:1992jt} that $SU(N)_k$ theory with one flavor experiences supersymmetry enhancement at the fixed point, therefore at the phase transition points we get $\mathcal{N}\,=\,2$ supersymmetry. The resulting phase diagram is summarised on fig. \eqref{PhaseDiag2}.

The next example we consider is $SU(3)$ theory with three fundamentals. In this case the baryon operator is cubic, and so this model can be considered as an $\mathcal{N}\,=\,1$ version of the \textit{critical scalar} (when written in components, \eqref{GeneralSuperpotential} produces quartic interaction for the scalars)\footnote{For a different version of $\mathcal{N}\,=\,1$ critical scalar see \cite{Benini:2018umh}.}. Our approach is to study the structure of IR phases in different regimes, and then sew the resulting patches together. More concretely, we consider semiclassical limits of large positive mass and large negative mass, behaviour near the point $m\,=\,0$, where the superpotential changes its asymptotic behaviour in some directions of field space, and also we are giving an effective description of the second order phase transition point, situating at some positive value of the mass. Quite remarkably, consistency of different patches suggests that at the phase transition point supersymmetry is enhanced to $\mathcal{N}\,=\,2$\footnote{There is plenty of literature devoted to supersymmetry enhancement in different instances, see e.g. \cite{Gaiotto:2018yjh, Benini:2018bhk, Bashmakov:2018ghn, Fazzi:2018rkr} for examples with two supercharges in the UV and e.g. \cite{Maruyoshi:2016tqk, Maruyoshi:2016aim, Agarwal:2016pjo, Benvenuti:2017lle, Benvenuti:2017kud, Agarwal:2017roi, Benvenuti:2017bpg, Agarwal:2018ejn, Gang:2018huc, Giacomelli:2018ziv, Carta:2018qke, Apruzzi:2018xkw, Agarwal:2018oxb, Aprile:2018oau, Garozzo:2019ejm, Carta:2019hbi} for examples with four or more supercharges in the UV.}. The resulting vacuum structure is depicted on fig. \eqref{SU3lambda}. 

The last example we consider is the case of $SU(4)$ gauge theory with four flavors. Anticipating the conclusion, we note that this example is sick, but in our opinion can serve as an illustration of the limitations that our arguments have.  We first perform the same type of analysis that was applied for the $SU(3)$ case, and again find that consistency requires SUSY enhancement. We then check this conjecture, performing the renormalization flow analysis and observe that supersymmetry enhancement can not really happen: the coupling $\lambda$ is IR free, and so there are no new IR stable $\mathcal{N}\,=\,2$ fixed points. We interpret this inconsistency as a consequence of the fact that the theory is not UV complete, exactly due to the IR freedom of the baryon coupling constant. Indeed, while working with an effective field theory, we can not reliably study vacua located at the large field values. Since information about them was crucial in our analysis of the phase structure, we can not really trust the conclusions following from it.

As a final remark, we would like to mention the $SU/U$ dualities for $\mathcal{N}\,=\,1$ SQCD discussed in \cite{Choi:2018ohn}. Following \cite{Aharony:2015mjs}, it is natural to assume that across this duality baryons on the $SU$ side are mapped to monopoles on the $U$ side. Our results then can be viewed as predictions for $\mathcal{N}\,=\,1$ $U(N)$ SQCD deformed by a monopole operator. It would be interesting to work out the picture directly from the $U$ side, providing a nontrivial check of the duality. We hope to report on it somewhere else.

The paper is organised in the following way. In the rest of this section we review the results about the phases of $SU(N)$ SQCD as a function of the mass parameter. In the three subsequent sections we describe our analysis of the $SU(N)$ QCD with $N$ flavors and baryon superpotential for $N\,=\,2,\ 3, \text{and}\ 4$.

\subsection{One-dimensional phase diagrams}\label{OneDimPhaseDiag}

Generally speaking, we are going to be interested in the infrared phases of $\mathcal{N}=1$ $G_k$ Chern-Simons theories (where $G$ denotes the gauge group and $k$ is the Chern-Simons (CS) level\footnote{Throughout this paper we will assume that $k\,>\,0$. The case of negative $k$ can be obtained by the time reversal transformation.}), interacting with some matter multiplets in a representation $\mathcal{R}$. It is straightforward to understand the IR dynamics when the matter multiplets are heavy, since they can be readily integrated out, leaving a pure vector multiplet in the infrared. Matter multiplets contain fermions, and upon their integration out CS level gets renormalized, with the renormalization being exact at one loop \cite{Coleman:1985zi}. E.g. integrating out one Dirac fermion in a representation $\mathcal{R}$, we get
\begin{equation}
    k \, \rightarrow \, k\,+\, \text{sign}(m)\,\times\, T(\mathcal{R}),
\end{equation}
with $T(\mathcal{R})$ being the index of the representation. Thus, we are going to get $G_{k\,-\,\frac{T(\mathcal{R})}{2}}$ theory for large negative mass and $G_{k\,+\,\frac{T(\mathcal{R})}{2}}$ for large positive mass. Below we will focus on the case of $SU(N)$ gauge group, which is going to be relevant for the following sections, but the general picture we state below holds for other gauge groups and representations as well.

Witten computed the index of $\mathcal{N}=1$ $SU(N)_k$ theories \cite{Witten:1999ds} (see also \cite{Smilga:2009ds}) with the result
\begin{equation}\label{IndexForSUN}
    \text{WI}(SU(N)_k)=
    \begin{cases}
    \ \binom{\tfrac{N}{2}\,+\,k\,-\,1}{N\,-\,1} & \text{for}\ k\,\geq\,\tfrac{N}{2}\\    
    \ \ \ \ \ \ \ 0    & \text{for}\ 0\,\leq\, k\,<\,\tfrac{N}{2}.
\end{cases}
\end{equation}
The result for $k\,\geq\,\tfrac{N}{2}$ can be understood as following. Supersymmetric completion of the Chern-Simons term is the mass term for the gaugini, and this mass is negative for positive $k$. Thus, at low enough energies we can integrate the gaugini out, generating the shift of the Chern-Simons level: $k\, \rightarrow\,k\,-\,\tfrac{N}{2}$. Also propagating gauge degrees of freedom are massive and can be integrated out, leaving the purely topological Chern-Simons theory $SU(N)_{k\,-\,\tfrac{N}{2}}$ in the IR. The index now corresponds to the number of states of this TFT put on a torus, or equivalently to the number of Wilson lines the theory possesses, which is indeed given by \eqref{IndexForSUN}. It is more involved to figure out the IR dynamics for the range $ 0<\, k\,<\,\tfrac{N}{2}$. In this case supesymmetry is dynamically broken, giving rise to a Majorana goldstino. Moreover, as it was argued recently in \cite{Gomis:2017ixy}, there is a decoupled TFT of the form
\begin{equation}
    U\left(\frac{N}{2}-k\right)_{\frac{N}{2}-k,N}.
\end{equation}

With \eqref{IndexForSUN} at our hands we immediately observe that the index for large and negative matter mass is different from that of the large and positive matter mass case. This can be explained, if we assume that while varying the mass from large negative to large positive values, at certain points new vacua come up from the field space infinity, making the index deficit up. In more general theories, we may expect co-dimension one \textit{walls} in the parameter space, on which the index jumps. A natural point for this to happen is the point $m\,=\,0$, where the superpotential changes its asymptotic behaviour. At this point the tree-level superpotential vanishes identically, therefore classically we have a moduli space of vacua, which is however expected to be affected by quantum corrections, allowed by the $\mathcal{N}=1$ supersymmetry. Since finally we expect a single vacuum, as the mass is increased this newly appeared vacua have to consequently merge with each other and with the old vacuum we had for the negative values of the mass. They can merge all together at once, or one by one, in a sequence of phase transitions. These phase transitions are going to be of the second order, since all the vacua are supersymmetric, and so have the same vanishing vacuum energy.

The discussion in the previous paragraph was rather schematic, but this picture has been worked out in full details for the cases of adjoint and fundamental matter in \cite{Bashmakov:2018wts}, \cite{Choi:2018ohn}. In particular, in \cite{Choi:2018ohn} the one-loop effective superpotential was computed\footnote{For the earlier computations of the one-loop effective superpotential for the case of matter in the adjoint representation see \cite{Armoni:2005sp}, \cite{Armoni:2006ee}.}, with the result
\begin{equation}\label{OneLoopEffPot}
    \mathcal{W}_{\text{one-loop}}\,=\, - \frac{\kappa}{8\pi}\,\text{Tr}\,\sqrt{\kappa^2\delta^{AB}\,+\,4g^2\,\bar{\Phi}^a\,T^{(A}T^{B)}\Phi_a},
\end{equation}
where $g$ is the gauge coupling, $\kappa=\tfrac{k\,g^2}{2\pi}$, $T^A$ is a generator of the gauge group in the representation $\mathcal{R}$, and $a$ is the flavor index.

We will now quote some results from \cite{Choi:2018ohn} that will be useful for us later on.

\begin{itemize}
    \item[1.] One of the simplest examples is the $SU(2)_k$ theory with one fundamental flavor. The large negative mass phase is given by $SU(2)_{k\,-\,\tfrac{1}{2}}$ vector multiplet, while the large positive mass phase is given by $SU(2)_{k\,+\,\tfrac{1}{2}}$ (note that $k$ is half-integer in this case). In the range of masses $0\,<\,m\,<\,m_*$ there is  additional trivial gapped vacuum. It comes from infinity as soon as we cross the point $m\,=\,0$, and merge with the $SU(2)_{k\,-\,\tfrac{1}{2}}$ vacuum at some point $m_*$, producing the $SU(2)_{k\,+\,\tfrac{1}{2}}$ vacuum. This is the second order phase transition point, and supersymmetry is enhanced at the CFT point to $\mathcal{N}\,=\,2$ \cite{Choi:2018ohn,Bashmakov:2018ghn}. The total Witten index is conserved: $(k\,-\,\tfrac{1}{2})\,+\,1\,=\,k\,+\,\tfrac{1}{2}$.
    \item[2.] In the case of $SU(2)_k$ gauge theory with two flavors the large negative mass phase corresponds to the $SU(2)_{k-1}$ vector multiplet with $\text{WI}= k$, while the large positive mass phase is given by  the $SU(2)_{k+1}$ vector multiplet with $\text{WI}=k+2$. At the point $m=0$ classically we have a moduli space of vacua. Note that this theory possesses not just the naive $U(2)$ global symmetry, but rather $Sp(2)$\footnote{See \cite{Choi:2018ohn} for a nice recent review.}. $Sp(2)$ multiplets are formed as $\tilde{\phi}^i_M\,=\,(\phi^i_1,\,\phi^i_2, \epsilon^{ij}\bar{\phi}^1_j,\epsilon^{ij}\bar{\phi}^2_j)$, and the gauge invariant description of the vacua can be given in terms of the meson matrix $\mathcal{M}^M_N\,=\,\bar{\tilde{\phi}}^M_i\, \tilde{\phi}^i_N$. Acting on $\tilde{\phi}$ by the gauge and global transformations, we can put $\mathcal{M}$ to the form
    \begin{equation}
    \left(
        \begin{matrix}
        v & 0 & 0 & 0 \\
        0 & 0 & 0 & 0 \\
        0 & 0 & v & 0 \\
        0 & 0 & 0 & 0
        \end{matrix}
        \right),
    \end{equation}
    having in this way a one dimensional classical moduli space. However, it is going to be lifted at the quantum level. Indeed, using the explicit form of \ref{OneLoopEffPot} and assuming small value of the mass parameter (such that we are still close to the classical moduli space point), we get
    \begin{equation}
        \mathcal{W}\,=\, m\,v^2 \, - \, \frac{3\kappa}{8\pi}\,\sqrt{\kappa^2+4\,g^2\,v^2}.
    \end{equation}
    It is easy to see that the equation $\mathcal{W}'\,=\,0$ for  $m\,\leq\,0$ has a single solution at the origin, and for $0\,<\,m\,<\tfrac{3g^2}{4\pi}$ has two solutions, at the origin and for a finite value of $v$. The solution at the origin is identified with the vacuum that we have had for large and negative mass, while a new vacuum appeared from the field space infinity breaks the global symmetry to $Sp(1)$, and is described at low energies by the non-linear sigma model (NLSM) with the target space
    \begin{equation}
        \frac{Sp(2)}{Sp(1)\,\times\,Sp(1)}\, \simeq \, S^2.
    \end{equation}
    The Witten index for a NLSM is given by the Euler characteristic of the target space manifold \cite{Hori:2003ic}, which for the present case is $2$. When $m\,>\tfrac{3g^2}{4\pi}$, the two vacua merge into a single vacuum through a second order phase transition. The index is again conserved across the phase transition: $(k\,-\,1)\,+\,2=\,k\,+\,1$
    \item[3.] \textbf{$SU(3)_k$ with three flavors.} For large negative mass at low energies we have $\mathcal{N}\,=\,1$ $SU(3)_{k\,-\,\tfrac{3}{2}}$ vector multiplet. When we cross the point $m\,=\,0$, three new vacua appear:
    \begin{eqnarray}
&\mathcal{N}\,=\,1&\ SU(2)_{k-1}\,\times \, \frac{U(3)}{U(2)\times U(1)}\ \ \text{NLSM},\ \ \text{WI}= (k-1) \times 3,\nonumber\\
&\mathcal{N}\,=\,1&\ \frac{U(3)}{U(2)\,\times\, U(1)}\ \  \text{NLSM},\ \ \text{WI}=3,\nonumber\\
&\mathcal{N}\,=\,1&\  S^1\ \ \text{NLSM},\ \ \text{WI}=0,\nonumber
\end{eqnarray}
Above $S^1$ corresponds to a Goldstone boson, associated with spontaneously broken $U(1)_B$. For some critical value $m\,=\,m_*$ all four vacua merge, such that for $m\,>\,m_*$ there is just one vacuum carrying $\mathcal{N}\,=\,1$ $SU(3)_{k\,+\,\tfrac{3}{2}}$ vector multiplet. We have\footnote{Here we use the following formula for the Euler characteristic of the coset space:
\begin{equation}
\chi\left(\frac{U(N)}{U(M)\,\times\,{U(N\,-\,M)}}\right)\,=\,\binom{N}{M}.    
\end{equation}
} $\frac{(k\,-\,1)(k\,-\,2)}{2}\,+\,3(k\,-\,1)\,+\,3\,+\,0\,=\,\frac{(k\,+\,1)(k\,+\,2)}{2}$.

\item[4.] \textbf{$SU(4)_k$ with four flavors.} For large negative mass at low energies we have $\mathcal{N}\,=\,1$ $SU(4)_{k\,-\,2}$ vector multiplet. When we cross the point $m\,=\,0$, four new vacua appear:
\begin{eqnarray}
&\mathcal{N}\,=\,1&\ SU(3)_{k\,-\,\tfrac{3}{2}}\,\times \, \frac{U(4)}{U(3)\times U(1)}\ \ \text{NLSM},\ \ \text{WI}= \frac{(k\,-\,1)(k\,-\,2)}{2} \times 4,\nonumber\\
&\mathcal{N}\,=\,1&\ SU(2)_{k\,-\,1}\,\times \, \frac{U(4)}{U(2)\times U(2)}\ \ \text{NLSM},\ \ \text{WI}= (k\,-\,1) \times 6,\nonumber\\
&\mathcal{N}\,=\,1&\ \frac{U(4)}{U(3)\,\times\, U(1)}\ \  \text{NLSM},\ \ \text{WI}=4,\nonumber\\
&\mathcal{N}\,=\,1&\  S^1\ \ \text{NLSM},\ \ \text{WI}=0.\nonumber
\end{eqnarray}
We can again check that $\frac{(k\,-\,1)(k\,-\,2)(k\,-\,3)}{6}\,+\,\frac{(k\,-\,1)(k\,-\,2)}{2}\,\times\,4\,+\,(k\,-\,1)\,\times\,6\,\,+\,4\,+\,0\,=\,\frac{(k\,+\,3)(k\,+\,2)(k\,+\,1)}{6}$.
\end{itemize}


\section{$SU(2)$ gauge group with two flavors}

In this section we will chart the phase diagram of $\mathcal{N}=1$ $SU(2)_k$ gauge theory with two flavors and with the superpotential
\begin{equation}
    \mathcal{W} = m\: \text{Tr}\: Q\bar{Q} + \lambda\, B + \bar{\lambda}\,\bar{B},
\end{equation}
where $B$ is the baryon operator, $B=\text{det}Q = \tfrac{1}{2}\, \epsilon^{ab}\,\epsilon_{ij} \,Q^i_a\, Q^j_b$, $(a,\, b)$ being flavor indices an $(i,\, j)$ being colour indices. Without loss of generality we will consider $\lambda$ to be real and positive (in this and in the following sections), even though sometimes it is useful to keep in mind that in principle it can take complex values.

\begin{figure}
    \centering
    \includegraphics{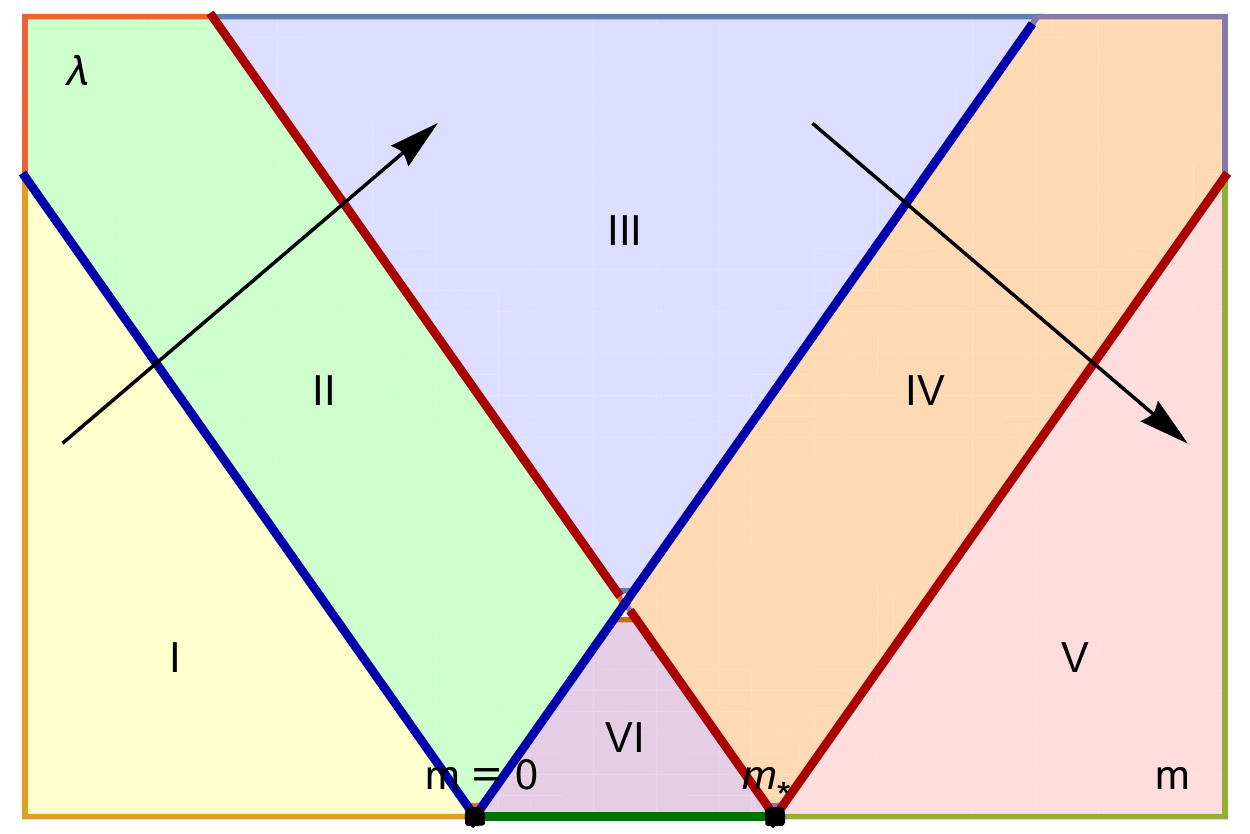}
    \caption{Phase diagram for $SU(2)_k$ theory with two flavors. In phase $\textrm{I}$ there is one vacuum, supporting $SU(2)_{k-1}$ vector multiplet. In phase $\textrm{II}$ there are two vacua: one supporting $SU(2)_{k-1}$ vector multiplet, and one trivial gapped. In phase $\textrm{III}$ there is one vacuum, supporting $SU(2)_{k}$ vector multiplet. In phase $\textrm{IV}$ there are two vacua: one supporting $SU(2)_{k}$ vector multiplet, and one trivial gapped. In phase $\textrm{V}$ there is one vacuum, supporting $SU(2)_{k+1}$ vector multiplet. Finally, in phase $\textrm{VI}$ there are three vacua: one supporting $SU(2)_{k-1}$ vector multiplet and two trivial gapped vacua. Blue lines correspond to the walls where the Witten index jumps, while red lines correspond to the second order phase transitions, which are described by two  $\mathcal{N}\,=\,2$ CFTs. Green line is the intermediate phase on the one-dimensional phase diagram with preserved $U(1)_B$.}
    \label{PhaseDiag2}
\end{figure}

Our proposal for the phase diagram of this theory as a function of $m$ and $\lambda$ is depicted on fig. \eqref{PhaseDiag2}. Below we will explain and motivate this proposal.

To start with, it is useful to understand the large mass limits, where the matter multiplets can be integrated out  semiclassically. Both superpotential couplings $m$ and $\lambda$ are mass parameters, and for their generic values we have two modes with masses $m\, -\, \lambda$ and $m\, +\, \lambda$. Thereby, we have three different regimes to consider.
\begin{itemize}
    \item When $m\, > \, \lambda$, and assuming $m,\:\lambda$ are large, we can integrate out two positive mass fermions in the fundamental representation, getting $SU(2)_{k+1}$ pure vector multiplet. This corresponds to the region $\textrm{V}$ on fig. \eqref{PhaseDiag2}. 
    \item When $-\lambda \, < \, m \, <\,\lambda$, there is a positive mass mode and a negative mass mode. When fermions are integrated out, we get $SU(2)_k$ pure vector multiplet. This phase corresponds to the region $\textrm{III}$ on fig. \eqref{PhaseDiag2}. 
    \item  Finally, when $m\, < \, -\lambda$, we have two negative mass modes, and integrating them out we get $SU(2)_{k-1}$ pure vector multiplet. This is the region $\textrm{I}$ on fig. \eqref{PhaseDiag2}.
\end{itemize}

Transitions between different semiclassical phases takes place near the locus $m\,=\,\pm\lambda$ of the parameter space: there one mode becomes massless, and classical moduli space of vacua appears. Therefore, quantum corrections become important for understanding of the dynamics, and more subtle analysis is required.

To proceed, consider first the transition between phase $\textrm{I}$ and phase $\textrm{III}$ along the line in the parameter space $m\,-\,\lambda\,=\,\text{const}$, and with $m\,-\,\lambda\,$ being large and negative: see the arrow in the upper-left corner of fig. \eqref{PhaseDiag2}. Along this trajectory one mode remains heavy, with the negative mass, while the second mass eigenvalue changes from large negative values, through zero, to large positive values. The heavy mode can be integrated out, and we get the $SU(2)_{k-1/2}$ theory with one fundamental flavor, whose phase diagram was reviewed in section \eqref{OneDimPhaseDiag}. In particular, phase $\textrm{I}$ corresponds to the negative mass phase of this theory, while phase $\textrm{III}$ corresponds to the large positive mass phase. We know that the transition between the two phases occurs in two steps. First, we cross the wall, and a new trivial gapped vacuum appears in addition to the topological vacuum we have had for negative mass. Second, the two vacua merge in a second order phase transition. On fig. \eqref{PhaseDiag2} the wall is denoted by the blue line, the intermediate phase with two vacua is phase $\textrm{II}$, and the phase transition is denoted by the red line. We have already mentioned that at the CFT point supersymmetry is enhanced to $\mathcal{N}=2$. We conjecture that the wall can be continued up to the point $m\,=\,\:\lambda\,=\,0$ (the wall of the one dimensional phase diagram), while the phase transition line can be continued up to the point $m\,=\,m_*,\: \lambda\,=\,0$ (the phase transition point of the one dimensional diagram).

Let us now move to the transition between phase $\textrm{III}$ and phase $\textrm{V}$, considering the line
$m\,+\,\lambda\,=\,\text{const}$, and with $m\,+\,\lambda$ being large and positive (see the arrow on the upper-right corner of diagram \eqref{PhaseDiag2}). Again, along this line one mass eigenvalue remains large, while the second eigenvalue changes its value from large negative, through zero, to large positive. Once again, the heavy mode can be integrated out, and the transition is described by the $SU(2)_{k+\tfrac{1}{2}}$ theory with one fundamental flavor. We cross the wall (the blue line), and the new trivial gapped vacuum comes up from infinity in addition to the negative mass topological vacuum: this is phase $\textrm{IV}$. Two vacua then merge in the second order phase transition (the red line), giving rise to another $\mathcal{N}\,=\,2$ CFT. Also here we assume that the wall can be continued up to the point $m\,=\,\:\lambda\,=\,0$, and the phase transition line can be continued up to the point $m\,=\,m_*,\: \lambda\,=\,0$. From these consideration we automatically get phase $\textrm{VI}$, consisting of the three vacua: on supporting $SU(2)_{k-1}$ vector multiplet and two trivial gapped vacua.

We will now provide two checks of the picture described above.

\textit{ 1). behaviour near the point $m\,=\,0,\: \lambda\,=\,0$ }

When both $m$ and $\lambda$ are small, we can trace the appearance of new vacua by first considering the classical moduli space, and observing its lifting due to the added mass terms and one-loop quantum corrections to the superpotential.

Classical moduli space can be described by the $2\times2$ matrix $q_a^i$, which can be put in the diagonal form by gauge and flavor rotations\footnote{Since the baryon deformation is not consistent with the $Sp(2)$ global symmetry discussed in the introduction, we have only $SU(2)_F$ at our disposal}:
\begin{equation}
    \left(
    \begin{matrix}
    q_1 & 0 \\
    0 & q_2 
    \end{matrix}
    \right),
\end{equation}
where $q_1$ can be made real. The superpotential for the moduli $q_i$ takes now the form
\begin{equation}
    \mathcal{W}\,=\,m\,(|q_1|^2+|q_2|^2)\,+\,\lambda (q_1\,q_2\,+\,\bar{q_1}\,\bar{q_2})\,-\,\frac{3\kappa}{8\pi}\sqrt{\kappa^2\,+\,4g^2\,(|q_1|^2\,+\,|q_2|^2)},
\end{equation}
with the $F$-term equations
\begin{align}
    \bar{\partial}_1\,\mathcal{W}\,=\,m\,q_1\,+\,\lambda\,\bar{q_2}\,-\,\frac{3\kappa}{8\pi}\,\frac{4g^2\,q_1}{\sqrt{\kappa^2\,+\,4g^2\,(|q_1|^2\,+\,|q_2|^2)}}\,=\,0,\nonumber\\
    \partial_2\,\mathcal{W}\,=\,m\,\bar{q}_2\,+\,\lambda\,q_1\,-\,\frac{3\kappa}{8\pi}\,\frac{4g^2\,\bar{q}_2}{\sqrt{\kappa^2\,+\,4g^2\,(|q_1|^2\,+\,|q_2|^2)}}\,=\,0.\nonumber
\end{align}
These equations have solutions away from the origin only if one of the compatibility conditions is satisfied:
\begin{equation}
    m\,-\,\frac{3\kappa}{8\pi}\,\frac{4g^2}{\sqrt{\kappa^2\,+\,4g^2\,(|q_1|^2\,+\,|q_2|^2)}}\,=\,\pm\lambda.
\end{equation}
When $m\,<\,-\lambda$, these conditions can not be solved, and so we have just one vacuum at the origin, with $q_1\,=\,q_2\,=\,0$. When $-\lambda\,<\,m\,<\,\lambda$, we can have
\begin{equation}
    \sqrt{\kappa^2\,+\,4g^2\,(|q_1|^2\,+\,|q_2|^2)}\,=\,(m\,+\,\lambda)\,\frac{2\pi\,g^2}{3\kappa},\; \; \; -\lambda\,<\,m\,<\,\lambda, \nonumber
\end{equation}
which gives one new solution of the full system of equation, and consequently one new trivial gapped vacuum. When $m\,>\,\lambda$, we can have
\begin{equation}
    \sqrt{\kappa^2\,+\,2g^2\,(|q_1|^2\,+\,|q_2|^2)}\,=\,(m\,\pm\,\lambda)\,\frac{4\pi\,g^2}{3\kappa},\; \; \; m\,>\,\lambda, \nonumber
\end{equation}
meaning that there are two new solutions of the full system, or two new trivial gapped vacua.

To summarise, we observe that while varying the mass parameter for a fixed small value of $\lambda$, we cross the wall twice, each time creating a new vacuum . This is in agreement with the phase diagram on fig. \eqref{PhaseDiag2}.

\textit{2). Behaviour near the point $m\,=\,m_*,\: \lambda\,=\,0$ }

We can also test our proposal considering effective description of the phase transitions. Assuming that only classically relevant and marginal operators are important, we consider the $SU(2)_k$ theory with the superpotential
\begin{align}
    \mathcal{W}\,=\, \tilde{m}\,\text{Tr}\,Q\bar{Q}\,+\,\lambda\,B\, +\,\bar{\lambda}\, \bar{B}\,+\,\frac{1}{2}\text{Tr}\,Q\bar{Q}Q\bar{Q}\,+\,\frac{\alpha}{2}\,\left(\text{Tr}\,Q\bar{Q}\right)^2\,+\nonumber\\
    +\,\frac{\beta}{2}\,B^2\, +\,\frac{\bar{\beta}}{2}\,\bar{B}^2\,+\,\gamma\,B\bar{B}, 
\end{align}
where $\tilde{m}\,=\,m\,-\,m_*$. To study the vacuum structure of this theory, we again put the matrix $q_m^i$ in the diagonal form, and try to find solutions of the resulting $F$-term equations:
\begin{align}
    \bar{\partial}_1\mathcal{W}\,=\,\tilde{m}\,q_1\,+\,\bar{\lambda}\bar{q_2}\,+q_1\,|q_1|^2\,+\,\alpha\,q_1\,(|q_1|^2\,+\,|q_2|^2)\,+\,\bar{q_2}\,(\bar{\beta}\,\bar{B}\,+\,\gamma\,B)\,=\,0,\nonumber\\
    \bar{\partial}_2\mathcal{W}\,=\,\tilde{m}\,q_2\,+\,\bar{\lambda}\bar{q_1}\,+q_2\,|q_2|^2\,+\,\alpha\,q_2\,(|q_1|^2\,+\,|q_2|^2)\,+\,\bar{q_1}\,(\bar{\beta}\,\bar{B}\,+\,\gamma\,B)\,=\,0.\nonumber
\end{align}
We always have a solution $q_1\,=\,q_2\,=\,0$. We observe further that we don't have solutions with one of the eigenvalues being zero, while another is different from zero. As such, let us concentrate on the solutions with both eigenvalues different from zero, and rewrite the equations in the following form, introducing a gauge-invariant meson matrix $M_m^n\,=\,q_m^i\bar{q}_i^n$:
\begin{equation}
    \tilde{m}\,M\,+\,\bar{\lambda}\,\bar{B}\,+\,M^2\,+\,\alpha\,M\,\text{Tr}\,M\,+\,\bar{B}\,(\bar{\beta}\,\bar{B}\,+\,\gamma\,B)=0.
\end{equation}
We see that both eigenvalues of $M$ satisfy the same equation, and so must be equal. We also note that $\text{det}M\,=\,B\,\bar{B}$. Assuming for simplicity that $\beta$ is real and positive (this will not change the net conclusion),  and assuming further that $1\,+\,2\alpha\,+\,\beta\,\,+\,\gamma>\,0$, we find the following structure of solutions.
\begin{itemize}
    \item For $\tilde{m}\,>\,\lambda$ the only solution is $M\,=\,0$, in this vacuum there are two positive fermions in the fundamental representation, and this corresponds to a single vacuum supporting the $SU(2)_{k+1}$ vector multiplet.
    \item  For $-\lambda\,<\,\tilde{m}\,<\,\lambda$ we get a solution
    \begin{equation}
        B\,=\,\frac{\tilde{m}\,-\,\lambda}{1\,+\,2\alpha\,+\,\beta\,+\,\gamma}.
    \end{equation}
    This gives a new trivial vacuum, while the vacuum in the origin contains $SU(2)_k$ vector multiplet (where we took into account that fermionic modes now have two opposite signs of masses).
    \item For $\tilde{m}\,<\,-\lambda$ we get two solutions
    \begin{equation}
        B\,=\,\frac{\pm\,\tilde{m}\,-\,\lambda}{1\,+\,2\alpha\,+\,\beta\,+\,\gamma}.
    \end{equation}
    Since now there are two negative mass fermions, we get a topological vacuum with $SU(2)_{k-1}$ vector multiplet at the origin, as well as two trivial gapped vacua.
\end{itemize}

As a result, we observe two phase transitions, both in agreement with the proposed phase diagram.

\section{$SU(3)$ gauge group with three flavors}\label{Section3}

We now turn to the consideration of $\mathcal{N}=1$ $SU(3)_k$ gauge theory with three flavors and once again in addition to the mass term add the baryon operator to the superpotential:
\begin{equation}
    \mathcal{W} = m\: \text{Tr}\: Q\bar{Q} + \lambda\, B + \lambda\,\bar{B},
\end{equation}
where this time $\,B=\,\text{det}\,Q = \tfrac{1}{3!}\, \epsilon^{abc}\,\epsilon_{ijk} \,Q^i_a\, Q^j_b\,Q^k_c$, $(a,\, b,\, c)$ being flavor indices an $(i,\, j,\, k)$ being colour indices.

We start with a study of the semiclassical picture, which is supposed to be valid when matter fields are heavy enough. $F$-term equations take the form

\begin{equation}\label{LargeMassFTerm}
    \frac{\partial\,\mathcal{W}}{\partial\,Q^i_{a}}\,=m\,\bar{Q}^{a}_i\,+\, \lambda\,\epsilon_{ijk}\,\epsilon^{abc}\, Q^{j}_{b}Q^{k}_{c}\,=\,0.
\end{equation}
For all values of the parameters we find the solution $Q\,=\,0$. If the matter field are heavy, we can integrate them out getting an $SU(3)_{k\,\pm\,\tfrac{3}{2}}$ vector multiplet in the IR, with the sign in the CS level depending on the sign of $m$.

In trying to find a solution away from the origin of the field space, we first note that if it exists, then $\text{rk}\,Q\,=\,3$. It allows us to rewrite \eqref{LargeMassFTerm} in the form
\begin{equation}
    m\,\bar{Q}\,+\,\lambda\,Q^{-1}\,\text{det}\,Q\,=\,0,
\end{equation}
or, introducing the meson matrix $M\,=\,Q_a^i\bar{Q}_i^b$, we get the equation
\begin{equation}\label{TreeLevelSU3}
    m\,M\,+\,\lambda\,B\,=\,0,
\end{equation}
which must be supplemented by the condition $\text{det}\,M\,=\,B\,\bar{B}$. We then get a new solution of the form
\begin{equation}
    M\,=\,\frac{m^2}{\lambda^2}\,\times\,I,\;\;\;\; B\,=\,-\frac{m^3}{\lambda^3},
\end{equation}
with $I$ being the $3\times3$ identity matrix. Semiclassically, when $m$ goes to zero, this solution approaches the origin, and the moduli space of vacua opens up.

\textit{ 1). Behaviour near the point $m\,=\,0,\: \lambda\,=\,0$ }

As the second step, we take quantum corrections into account and consider the behaviour of the theory near the wall, namely near $m\,=\,0$. When $m\,=\,0$, the classical moduli space is lifted, remaining the vacuum in the origin of the field space. Yet, for $\lambda\,\neq\,0$ another vacuum can be found. To see this, let us examine if there are vacua preserving the global symmetry. The strategy is to analyse the $F$-term equations for classical moduli. The matrix $Q^i_a$ can be put in the diagonal form with the aid of gauge and global symmetry rotations,
\begin{equation}
    Q^i_a\!=\!\left(
    \begin{matrix}
    q_1 && 0 && 0\\
    0 && q_2 && 0\\
    0 && 0 && q_3,
    \end{matrix}
    \right),
\end{equation}
and so the moduli can be chosen as $q_1,\, q_2,\, q_3$. General $F$-term equations for them take the form
\begin{equation}
    \bar{\partial}_i\mathcal{W}\,=\,\lambda\frac{\bar{q}_1\,\bar{q}_2\,\bar{q}_3}{\bar{q}_i}\,+\, \bar{\partial}_i\mathcal{W}_{\text{one-loop}}.
\end{equation}
The function $\mathcal{W}_{\text{one-loop}}$, though can be computed explicitly, has rather complicated form. However, it simplifies when we restrict ourselves to the case $q_1\!=\!q_2\!=\! q_3\!=\!q$, in which case the equation to be solved is
\begin{equation}
    \lambda\,\bar{q}^2\,-\,\frac{\kappa}{8\pi}\,\frac{16g^2\,q}{3\,\sqrt{\kappa^2\,+\,4g^2\,|q|^2}}\,=\,0.
\end{equation}
It always has a solution
\begin{equation}
    q\,=\,\frac{\sqrt{2}\kappa}{4g}\,\sqrt{-1\,+\,\sqrt{1\!+\!\frac{64g^4}{9\pi^2\lambda^2}}},
\end{equation}
which we identify with the trivial gapped vacuum we have had in the large negative mass limit. Classically it collides with the vacuum at the origin when $m$ goes to zero, but quantum corrections make two vacua to repel.

When the mass becomes positive, some other solutions appear. In particular, the solution with $q_1\,\neq\,0$, and $q_2\,=\,q_3\,=\,0$, existing for $\lambda\,=\,0$, is not affected by the baryon deformation. Its low energy dynamics is still given by
\begin{equation}
    \mathcal{N}\,=\,1\ \ \ SU(2)_{k-1} \ \times \ \frac{U(3)}{U(2)\times U(1)}\ \ \text{NLSM}.
\end{equation}

Differently from the $\lambda\,=\,0$ case, we do not have a solution with $q_1\,=\,q_2,\quad q_3\,=\,0$, but we have a solution of the form $q_1\,=\,q_2,\quad q_3\,=\,\mathcal{O}(\lambda)$, which also breaks the whole gauge group and has the same pattern of global symmetry breaking: $SU(3)\,\rightarrow\,S[U(2)\,\times\,U(1)]$. In the infrared this vacuum is described by
\begin{equation}
    \mathcal{N}\,=\,1\ \ \ \frac{U(3)}{U(2)\times U(1)}\ \ \text{NLSM}.
\end{equation}

We see that crossing the wall at $m\,=\,0$, we find a phase with four vacua. We don't have a proof that there are no other supersymmetric vacua, but the vacua we have described are enough to match the index of the large positive phase:
\begin{align}
&\text{WI}(SU(3)_{k-\tfrac{3}{2}})\,+\,\text{WI}(\text{trivial vacuum})\,+\nonumber\\
+&\,\text{WI}(SU(2)_{k-1} \ \times \ \frac{U(3)}{U(2)\times U(1)})\,+\,\text{WI}(\frac{U(3)}{U(2)\times U(1)})\,=\nonumber\\
=&\,\frac{(k-1)(k-2)}{2}\,+\,1\,+\,3\,(k-1)\,+\,3\,=\nonumber\\
=&\,\frac{(k+2)(k+1)}{2}\,+\,1\,=\,\text{WI}(SU(3)_{k+\tfrac{3}{2}})\,+\,\text{WI}(\text{trivial vacuum}).
\end{align}
These four vacua then must undergo a sequence of second order phase transitions, turning into the vacua of the large positive mass phase.

\begin{figure}
\centering
\begin{subfigure}{.4\textwidth}
  \centering
  \includegraphics[width=.9\linewidth]{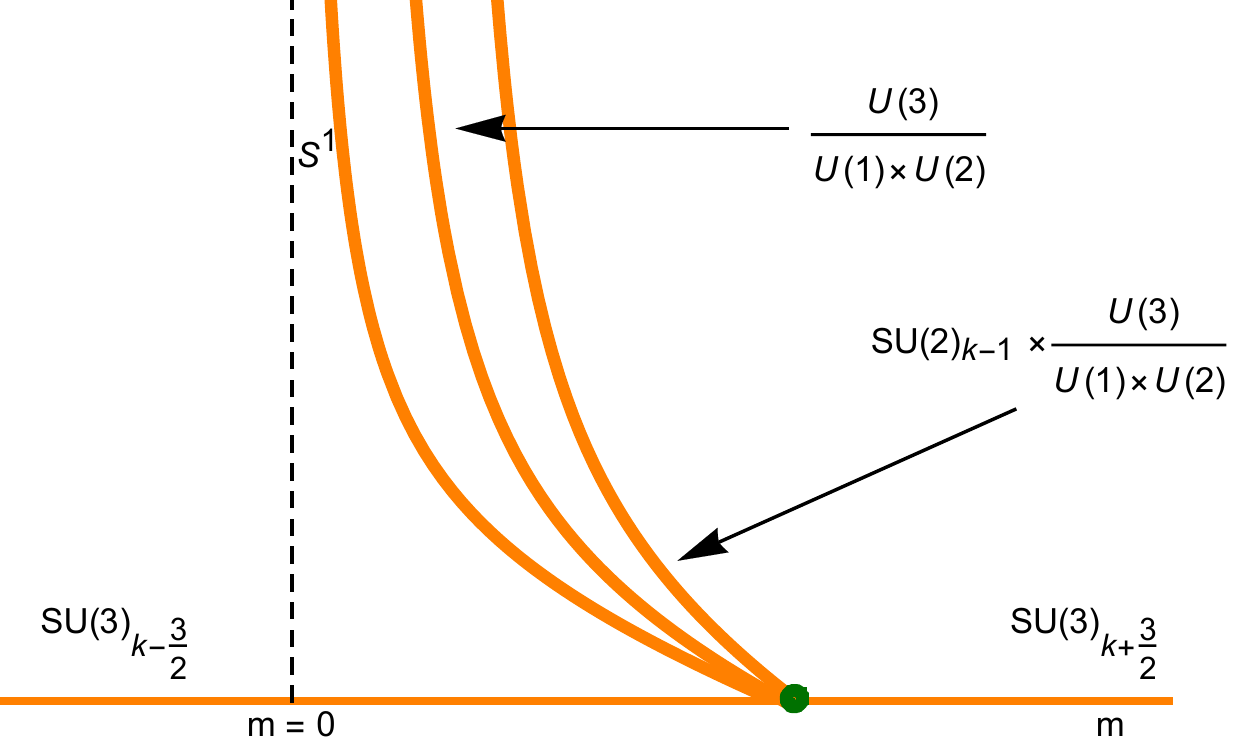}
  \caption{ }
  \label{SU3}
\end{subfigure}%
\begin{subfigure}{.4\textwidth}
  \centering
  \includegraphics[width=.9\linewidth]{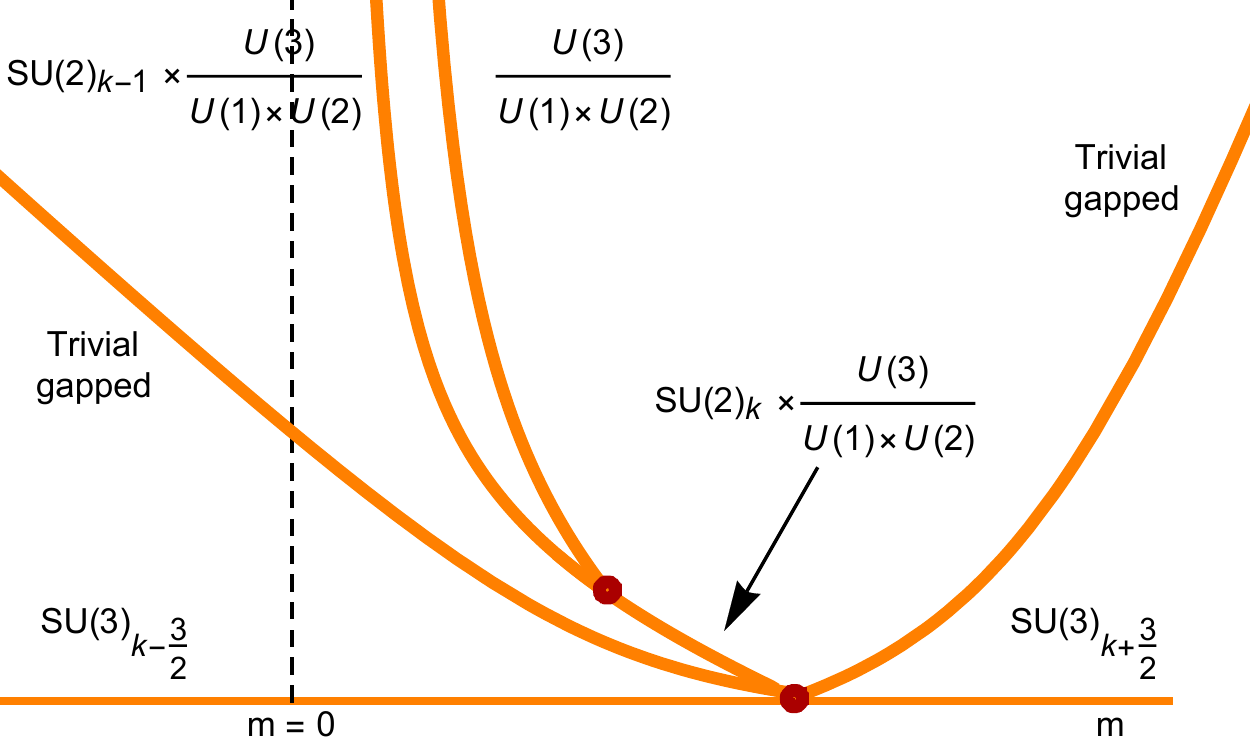}
  \caption{ }
  \label{SU3lambda}
\end{subfigure}
\caption{Schematic depiction of the vacuum structure of the $\mathcal{N}\,=\,1$ $SU(3)_k$ theory with three fundamental flavors. (a) without baryon deformation ($\lambda\,=\,0$). (b) with baryon deformation ($\lambda\,\neq\,0$). Green marker denotes the second order phase transition with $\mathcal{N}\,=\,1$ CFT. Red markers denote the second order phase transitions with $\mathcal{N}\,=\,2$ CFTs. }
\label{fig:test}
\end{figure}

\textit{2). Behaviour near the phase transition points }

In order to get more insights about the details of phase transitions points, we now use the effective description of the phase transition, assuming that classically relevant and marginal operators are the most important ones. We consider the $SU(3)_k$ theory coupled to three fundamental matter multiplets with the superpotential
\begin{equation}
    \mathcal{W}\,=\,\tilde{m}\,\text{Tr}\,Q\bar{Q}\,+\,\tfrac{1}{2}\,\text{Tr}\,Q\bar{Q}Q\bar{Q}\,+\,\tfrac{\alpha}{2}\,(\text{Tr}\,Q\bar{Q})^2\,+\,\lambda(B\,+\,\bar{B}),
\end{equation}
where again $\tilde{m}$ is the effective mass near the transition point. $F$-term equations take the form
\begin{equation}\label{EffectiveFTerm}
    \frac{\partial\,\mathcal{W}}{\partial\,\bar{Q}_i^{a}}\,=\tilde{m}\,Q^{i}_a\,+\,(Q\bar{Q}Q)^{i}_a\,+\,\alpha\,Q^{i}_a\,\text{Tr}(Q\bar{Q})\,+\,\lambda\,\epsilon^{ijk}\,\epsilon_{abc}\, \bar{Q}_{j}^{b}\bar{Q}_{k}^{c}\,=\,0.
\end{equation}
Gauge invariant solutions can be parametrized by the meson matrix $M_a^b\,=\,Q^{i}_a\,\bar{Q}_i^b$, and by the baryon operator $B$.

For the large positive mass, we expect to have two vacua: one at the origin, supporting the $SU(3)_{k+3}$ vector multiplet, and one trivial gapped vacuum, preserving $SU(3)_F$ symmetry. It is obvious that the first solution, corresponding to $M\,=\,B\,=\,0$, do exist, so let us check if we have the second solution, for which all the eigenvalues of $M$ are non-vanishing and equal to each other. Eq. \eqref{EffectiveFTerm} can be rewritten in the form
\begin{equation}
    \tilde{m}\,M\,+\,M^2\,+\,\alpha\,M\,\text{Tr}\,M\,+\,\lambda\,\bar{B}\,=\,0.
\end{equation}
This must be supplemented with the relation $\text{det}\,M\,=\,B\,\bar{B}$.

Straightforward analysis of this equation reveals the following picture. When $1\,+\,3\alpha\,>\,0$, and the mass is sufficiently large ($\tilde{m}\,>\,\frac{\lambda^2}{4(1\,+\,3\alpha)}$), there are no solutions of the kind described above, when $0\,<\,\tilde{m}\,<\,\frac{\lambda^2}{4(1\,+\,3\alpha)}$ there are two of them, 
\begin{equation}
    M\,=\,\left(\frac{\lambda\,\pm\,\sqrt{\lambda^2\,-\,4\tilde{m}(1\,+\,3\alpha)}}{2(1\,+\,3\alpha)}\right)^2\,\times\,I,\ \ \ B\,=\,-\left(\frac{\lambda\,\pm\,\sqrt{\lambda^2\,-\,4\tilde{m}(1\,+\,3\alpha)}}{2(1\,+\,3\alpha)}\right)^3,
\end{equation}
and when $\tilde{m}\,<\,0$ there are again two solutions:
\begin{equation}
    M\,=\,\left(\frac{\pm\,\lambda\,+\,\sqrt{\lambda^2\,-\,4\tilde{m}(1\,+\,3\alpha)}}{2(1\,+\,3\alpha)}\right)^2\,\times\,I,\ \ \ B\,=\,\mp\left(\frac{\pm\lambda\,+\,\sqrt{\lambda^2\,-\,4\tilde{m}(1\,+\,3\alpha)}}{2(1\,+\,3\alpha)}\right)^3.
\end{equation}
If, on the other hand, $1\,+\,3\alpha\,<\,0$, there are two solutions for positive mass, 
\begin{equation}
    M\,=\,\left(\frac{\pm\,\lambda\,+\,\sqrt{\lambda^2\,-\,4\tilde{m}(1\,+\,3\alpha)}}{2(1\,+\,3\alpha)}\right)^2\,\times\,I,\ \ \ B\,=\,\mp\left(\frac{\pm\lambda\,+\,\sqrt{\lambda^2\,-\,4\tilde{m}(1\,+\,3\alpha)}}{2(1\,+\,3\alpha)}\right)^3,
\end{equation}
two solutions for $\frac{\lambda^2}{4(1\,+\,3\alpha)}\,<\,\tilde{m}\,<\,0$,
\begin{equation}
    M\,=\,\left(\frac{\lambda\,\pm\,\sqrt{\lambda^2\,-\,4\tilde{m}(1\,+\,3\alpha)}}{2(1\,+\,3\alpha)}\right)^2\,\times\,I,\ \ \ B\,=\,-\left(\frac{\lambda\,\pm\,\sqrt{\lambda^2\,-\,4\tilde{m}(1\,+\,3\alpha)}}{2(1\,+\,3\alpha)}\right)^3,
\end{equation}
and no solutions for $\tilde{m}\,<\,\frac{\lambda^2}{4(1\,+\,3\alpha)}$.

This disagrees with our expectations, according to which we must see one $SU(3)_F$-preserving vavuum to the right from the phase transition points and one to the left from the phase transition points. A way out is to require that $1\,+\,3\alpha\,=\,0$. Indeed, having implemented this condition, we get the following equation for the eigenvalue $\mu$ of the meson matrix:
\begin{align}
    \tilde{m}\,\mu\,+\,\lambda\,\bar{B}\,=\,0,\nonumber\\
    B\,\bar{B}\,=\,\mu^3.
\end{align}
This is the same equation as eq. \eqref{TreeLevelSU3}, and it exactly reproduces the behaviour we expect.

The condition $1\,+\,3\alpha\,=\,0$ may look like inappropriate fine tuning, but in fact it becomes natural if we assume that supersymmetry at the fixed point is enhanced to $\mathcal{N}\,=\,2$: then $\alpha\,=\,-\frac{1}{3}$ is imposed by supersymmetry. The discussion above should not be considered as a proof of supersymmetry enhancement, but rather like an indication on that. Indeed, it is based on the assumption that irrelevant operators are not important for the description of this phase transition (or in other words, there are no dangerous irrelevant operators).

We now turn to the vacua breaking $SU(3)_F$, and corresponding to the meson matrix with two equal eigenvalues and one eigenvalue different from the first two. When $\tilde{m}\,<\,0$, a new solution appears:
\begin{equation}
    M\,=\,\left(
    \begin{matrix}
    -\tfrac{3}{2}\,\tilde{m} && 0 && 0 \\
    0 && 0 && 0\\
    0 && 0 && 0
    \end{matrix}
    \right), \: \: \: \: \:\: B\,=\,0.
\end{equation}
This vacuum breaks flavor symmetry as $SU(3)\,\rightarrow\,S[U(2)\,\times\,U(1)]$, and so we get a NLSM
\begin{equation}
    \frac{U(3)}{U(2)\,\times\,U(1)}
\end{equation}
in the IR. Moreover, gauge group is broken to $SU(2)$. If $-\tfrac{2}{3}\,\lambda^2\,<\,\tilde{m}\,<0$ we have two positive mass modes and two negative mass modes, so CS level does not get renormalized. If, on the other hands, $\tilde{m}\,<\,-\tfrac{2}{3}\,\lambda^2$, there is one positive mass mode and three negative mass modes, therefore the level gets shifted: $k\,\rightarrow\,k\,-\,1$. The resulting low-energy dynamics is provided by the following theories:
\begin{align}
    \mathcal{N}\,&=\!1\:\:\:SU(2)_k\:\times\:\text{NLSM}\ \ \frac{U(3)}{U(2)\,\times\,U(1)},\ \  -\tfrac{2}{3}\,\lambda^2\,<\,\tilde{m}\,<\,0,\nonumber\\
    \mathcal{N}\,&=\!1\:\:\:SU(2)_{k-1}\:\times\:\text{NLSM}\ \ \frac{U(3)}{U(2)\,\times\,U(1)},\ \  \tilde{m}\,<\,-\tfrac{2}{3}\,\lambda^2.\nonumber
\end{align}
When $\tilde{m}\,<\,-\tfrac{2}{3}\,\lambda^2$, another solution branches off from the one described above:
\begin{equation}
M\,=\,\left(
    \begin{matrix}
    \lambda^2 && 0 && 0 \\
    0 && -3\tilde{m}\,-\,2\lambda^2 && 0\\
    0 && 0 && -3\tilde{m}\,-\,2\lambda^2
    \end{matrix}
    \right), \: \: \: \: \:\: B\,=\,-(3\tilde{m}\,+\,2\lambda^2)\lambda.
\end{equation}
It breaks the gauge group completely, while the global symmetry is broken as $SU(3)\,\rightarrow\,S[U(2)\,\times\,U(1)]$, which corresponds to
\begin{equation}
    \mathcal{N}\,=\,1\:\:\:\text{NLSM}\ \ \frac{U(3)}{U(2)\,\times\!U(1)}
\end{equation}
in the infrared.

\vspace{10mm}

All our findings can be sewed together in a complete picture, summarised on fig. \eqref{SU3lambda}. For large negative mass we have two vacua, one preserving gauge symmetry and one trivial gapped vacuum. After crossing the wall at $m\,=\,0$, we find two new vacua, coming in from infinity. At some critical mass value these two vacua undergo a phase transition, merging into a single vacuum. The CFT at the fixed point is the same as for $SU(2)_{k\,-\,\tfrac{1}{2}}$ with one fundamental multiplet, and thus has $\mathcal{N}\,=\,2$ supersymmetry. Then the remaining three vacua experience another phase transition, after which we find the large positive mass semiclassical phase. We conjecture that around this fixed point the theory is dual to $\mathcal{N}\,=\,2$ $SU(3)_{k\,+\,\tfrac{3}{2}}$ theory with three chiral multiplets $\Phi^m$ ($m\,=\,1,\,2,\,3$), and with the superpotential
\begin{equation}
    \mathcal{W}_{\mathcal{N}\,=\,2}\,=\,\tfrac{1}{3!}\lambda\,\text{det}\Phi.
\end{equation}


\section{$SU(4)$ gauge group with four flavors}\label{Section4}
We finally move to our last example, namely $SU(4)_k$ theory with four flavors and with the superpotential
\begin{equation}
    \mathcal{W}\,=\,m\,\text{Tr}\,Q\,\bar{Q}\,+ \lambda\,(B\,+\,\bar{B}),
\end{equation}
where $\,B=\,\text{det}\,Q = \tfrac{1}{4!}\, \epsilon^{abcd}\,\epsilon_{ijkl} \,Q^i_a\, Q^j_b\,Q^k_c\,Q^l_d$, $(a,\, b,\, c,\, d)$ being flavor indices and $(i,\, j,\, k,\, l)$ being colour indices. In certain respects the situation here is similar to the $SU(3)$ case. Following the same line of steps, we first consider semiclassical phases at large masses, positive or negative. $F$-term equations are
\begin{equation}\label{LargeMassFTermSU4}
    \frac{\partial\,\mathcal{W}}{\partial\,Q^i_{a}}\,=m\,\bar{Q}^{a}_i\,+\, \lambda\,\epsilon_{ijkl}\,\epsilon^{abcd}\, Q^{j}_{b}Q^{k}_{c}Q^{l}_{d}\,=\,0.
\end{equation}
Again, we observe that there is a vacuum at the origin, $Q\,=\,0$, and any solution away from the origin must satisfy the condition $\text{rk}\,Q\,=\,4$. Rewriting the $F$-term equations in matrix form, we get
\begin{equation}
    m\,M\,+\,\lambda\, B\,=\,0,
\end{equation}
together with the relation $\text{det}\,M\,=\,B\,\bar{B}$. It is then easy to obtain the solution we are looking for:
\begin{equation}
    M\,=\,\frac{|m|}{\lambda}\,\times\,I,\;\;\;\; B\,=\,-\text{sign}(m)\frac{m^2}{\lambda^2}.
\end{equation}
As in the previous case, in the semiclassical phases we have a trivial gapped vacuum in addition to the topological vacuum in the origin. We will now discuss the behaviour of vacuum solutions for small $m$ and $\lambda$.

\textit{ 1). Behaviour near the point $m\,=\,0,\: \lambda\,=\,0$.}

Vacuum structure in this regime can be extracted from the superpotential composed out of the mass deformation, baryon deformation, and one-loop quantum corrections:
\begin{equation}
    \mathcal{W}\,=\,m\,\text{Tr}\,Q\bar{Q}\,+\,\lambda\,(B\,+\,\bar{B})\,+\,\mathcal{W}_{\text{one-loop}}.
\end{equation}
As before, we start with the classical moduli space, parametrized by the matrix $Q^i_m$ which can be brought to the diagonal form by gauge and global symmetry rotations: $Q\,=\,(q_1,\, q_2,\, q_3,\, q_4)$. The superpotential above should be written for these moduli, and determines how the moduli space is lifted. While the one-loop contribution is represented by a fairly complicated function, its form simplifies when we are looking for solutions with $q_1\,=\,q_2\,=\,q_3\,=\,q_4\,=\,q$. For this case the $F$-term equation to consider takes the form (we put $m\,=\,0$ for a moment)
\begin{equation}\label{FtermSU4}
    \lambda\,\bar{q}^3\,-\,\frac{15\kappa}{16\pi}\,\frac{g^2\,\bar{q}}{\sqrt{\kappa^2\,+\,4g^2\,|q|^2}}\,=\,0,
\end{equation}
which always has a solution \footnote{It is easy to see that the problem is equivalent to the search of a positive root of a cubic equation. The cubic equation is of a special kind, and always allows for a solution.}. This symmetry preserving vacuum should be identified with the trivial vacuum we found in the large negative mass limit.

When $m\,>\,0$, eq. \eqref{FtermSU4} also admits solutions with $q_1\,\neq\,0$, $q_2\,=\,q_3\,=\,q_4\,=\,0$ and with $q_1\,=\,q_2\,\neq\,0$, $q_3\,=\,q_4\,=\,0$. These solutions are the same as for $\lambda\,=\,0$, and at low energies reduce to
\begin{equation}
\mathcal{N}\!=\!1\:\:\:SU(3)_{k\,-\,\tfrac{3}{2}}\:\times\:\text{NLSM}\ \ \frac{U(4)}{U(3)\!\times\!U(1)},\ \ \text{WI}\,=\,\frac{(k-1)(k-2)}{2}\,\times\,4
\end{equation}
and
\begin{equation}
    \mathcal{N}\!=\!1\:\:\:SU(2)_{k\,-\,1}\:\times\:\text{NLSM}\ \ \frac{U(4)}{U(2)\!\times\!U(2)},\ \ \text{WI}\,=\,(k\,-\,1)\,\times\,4,\qquad\ \ 
\end{equation}
respectively.

We don't have a solution with three eigenvalues equal to each other while the forth is zero, but for $m\,>\,0$ there is a solution of the form $q_1\,=\,q_2\,=\,q_3$, $q_4\,=\,\mathcal{O}(\lambda)$. It flows to
\begin{equation}
    \mathcal{N}\!=\!1\:\:\:\text{NLSM}\ \ \frac{U(4)}{U(3)\!\times\!U(1)},\ \ \text{WI}\,=\,4
\end{equation}
in the infrared. In summary, when we the mass is small and positive, we still have two vacua we have had for large and negative masses, and in addition three more vacua come from infinity. We expect that due to certain second order phase transitions these vacue will turn into the two vacua we have found for the large positive mass limit. It is easy to check that the total Witten index is indeed the one expected for the large positive mass phase.

\textit{ 2). Behaviour near the phase transition points.}

To proceed, we consider the effective description of the phase transition point(s), considering the $SU(4)_k$ theory with the superpotential
\begin{equation}
    \mathcal{W}\,=\, \tilde{m}\,\text{Tr}\,Q\bar{Q}\,+\,\tfrac{1}{2}\,\text{Tr}\,Q\bar{Q}Q\bar{Q}\,+\,\tfrac{\alpha}{2}\,(\text{Tr}\,Q\bar{Q})^2\,+\,\lambda\,(B\,+\,\bar{B}),
\end{equation}
with $\tilde{m}\,=\,m\,-\,m_*$. The $F$-term equations following from the superpotential above are given by
\begin{equation}\label{LargeMassFTermSU4CFT}
    \frac{\partial\,\mathcal{W}}{\partial\,Q^i_{a}}\,=\tilde{m}\,\bar{Q}^{a}_i\,+\,(\bar{Q}Q\bar{Q})^a_i\,+\,\alpha\,\bar{Q}^a_i\,\text{Tr}Q\bar{Q}\,+ \lambda\,\epsilon_{ijkl}\,\epsilon^{abcd}\, Q^{j}_{b}Q^{k}_{c}Q^{l}_{d}\,=\,0.
\end{equation}
Since we expect to have a trivial gapped vacuum for positive mass values, we start by looking for a solution, preserving $SU(4)_F$. The $F$-term equations then can be rewritten in terms of the meson matrix $M$ defined as above:
\begin{equation}
    \tilde{m}\,M\,+\,M^2\,+\,\alpha\,M\,\text{Tr} M\,+\,\lambda\,\bar{B}\,=\,0
\end{equation}
together with the relation $\text{det}\,M\,=\,B\bar{B}$.
\begin{center}
\begin{tabular}{ |c|c|c| } 
 \hline
 $1\,+\,4\alpha\,>\,0$ & $0\,<\,\lambda\,<\,1\,+\,4\alpha$ & $\lambda\,>\,1+4\alpha$ \\ 
 \hline
 $\tilde{m}\,>\,0$ & $\emptyset$ & $-\frac{\tilde{m}}{1\,+4\alpha\,-\,\lambda}$ \\[1ex]
 \hline
 $\tilde{m}\,<\,0$  & $-\frac{\tilde{m}}{1\,+4\alpha\,+\,\lambda}\;\;\&\;\; -\frac{\tilde{m}}{1\,+4\alpha\,-\,\lambda}$ & $-\frac{\tilde{m}}{1\,+4\alpha\,+\,\lambda}$ \\[1ex]
 \hline\hline
 $1\,+\,4\alpha\,<\,0$ & $0\,<\,\lambda\,<-\,1\,-\,4\alpha$ & $\lambda\,>\,-1-4\alpha$ \\ 
 \hline
 $\tilde{m}\,>\,0$ & $-\frac{\tilde{m}}{1\,+4\alpha\,+\,\lambda}\;\;\&\;\; -\frac{\tilde{m}}{1\,+4\alpha\,-\,\lambda}$ & $-\frac{\tilde{m}}{1\,+4\alpha\,-\,\lambda}$ \\[1ex]
 \hline
 $\tilde{m}\,<\,0$  & $\emptyset$ & $-\frac{\tilde{m}}{1\,+4\alpha\,+\,\lambda}$ \\[1ex]
 \hline
\end{tabular}
\label{tableSU4}
\end{center}

The number of solutions for different values of $\alpha$, $m$ and $\lambda$ can be read off from table \eqref{tableSU4}. When $1\,+\,4\alpha\,>\,0$ and $0\,<\,\lambda\,<\,1\,+\,4\alpha$, there are no solutions preserving $SU(4)_F$  and breaking $SU(4)_{\text{gauge}}$ for positive mass, and two of the for negative mass. On the other hand, when $1\,+\,4\alpha\,<\,0$ and $0\,<\,\lambda\,<\,-1\,-\,4\alpha$ there are two such solutions for positive mass and no solutions for negative mass. This is in contrast with our expectations, since we expect to find one (and only one) solution of this kind for any $\lambda$. The way out is to assume that actually $1\,+\,4\alpha\,=\,0$. This apparent fine tuning can again be explained by the assumption that at the fixed point supersymmetry is enhanced to $\mathcal{N}\,=\,2$. Let us assume that this indeed happens, and $\alpha\,=\,-\tfrac{1}{4}$.

Let us list the other vacua appearing for $\tilde{m}\,<\,0$. The solution
\begin{equation}
M\,=\,\left(
\begin{matrix}
-\tfrac{4}{3}\,\tilde{m} && 0 && 0 && 0 \\
0 && 0 && 0 && 0 \\
0 && 0 && 0 && 0 \\
0 && 0 && 0 && 0 
\end{matrix}
\right),\:\:\:\: B\,=\,0
\end{equation}
breaks global symmetry as $SU(4)\,\rightarrow\,S[U(3)\,\times\,U(1)]$ and gauge symmetry as
$SU(4)\,\rightarrow\,SU(3)$. It is described by
\begin{equation}
    \mathcal{N}\,=\,1\:\:\:SU(3)_{k\,-\,\tfrac{3}{2}}\,\times\,\frac{U(4)}{U(3)\,\times\,U(1)}\,\,\text{NLSM}.
\end{equation}
The solution
\begin{equation}
M\,=\,\left(
\begin{matrix}
-2\,\tilde{m} && 0 && 0 && 0 \\
0 && -2\,\tilde{m} && 0 && 0 \\
0 && 0 && 0 && 0 \\
0 && 0 && 0 && 0 
\end{matrix}
\right),\:\:\:\: B\,=\,0
\end{equation}
breaks global symmetry as $SU(4)\,\rightarrow\,S[U(2)\,\times\,U(2)]$ and gauge symmetry as
$SU(4)\,\rightarrow\,SU(2)$. It is described by
\begin{equation}
    \mathcal{N}\,=\,1\:\:\:SU(2)_{k\,-\,1}\,\times\,\frac{U(4)}{U(2)\,\times\,U(2)}\,\,\text{NLSM}.
\end{equation}
Finally, the solution
\begin{equation}
    M\,=\,\left(
    \begin{matrix}
    -\frac{4\tilde{m}}{1\,+\,3\lambda^2} && 0 && 0 && 0\\
    0 && -\frac{4\tilde{m}}{1\,+\,3\lambda^2} && 0 && 0\\
    0 && 0 && -\frac{4\tilde{m}}{1\,+\,3\lambda^2} && 0\\
    0 && 0 && 0 && -\frac{4\tilde{m}\,\lambda^2}{1\,+\,3\lambda^2}
    \end{matrix}
    \right),\:\:\:B\,=\,\frac{16\tilde{m}^2\,\lambda}{(1\,+\,3\lambda^2)^2}.
\end{equation}

\subsection{Tensions with the RG flow analysis}

\begin{figure}
    \centering
    \includegraphics[width=.5\linewidth]{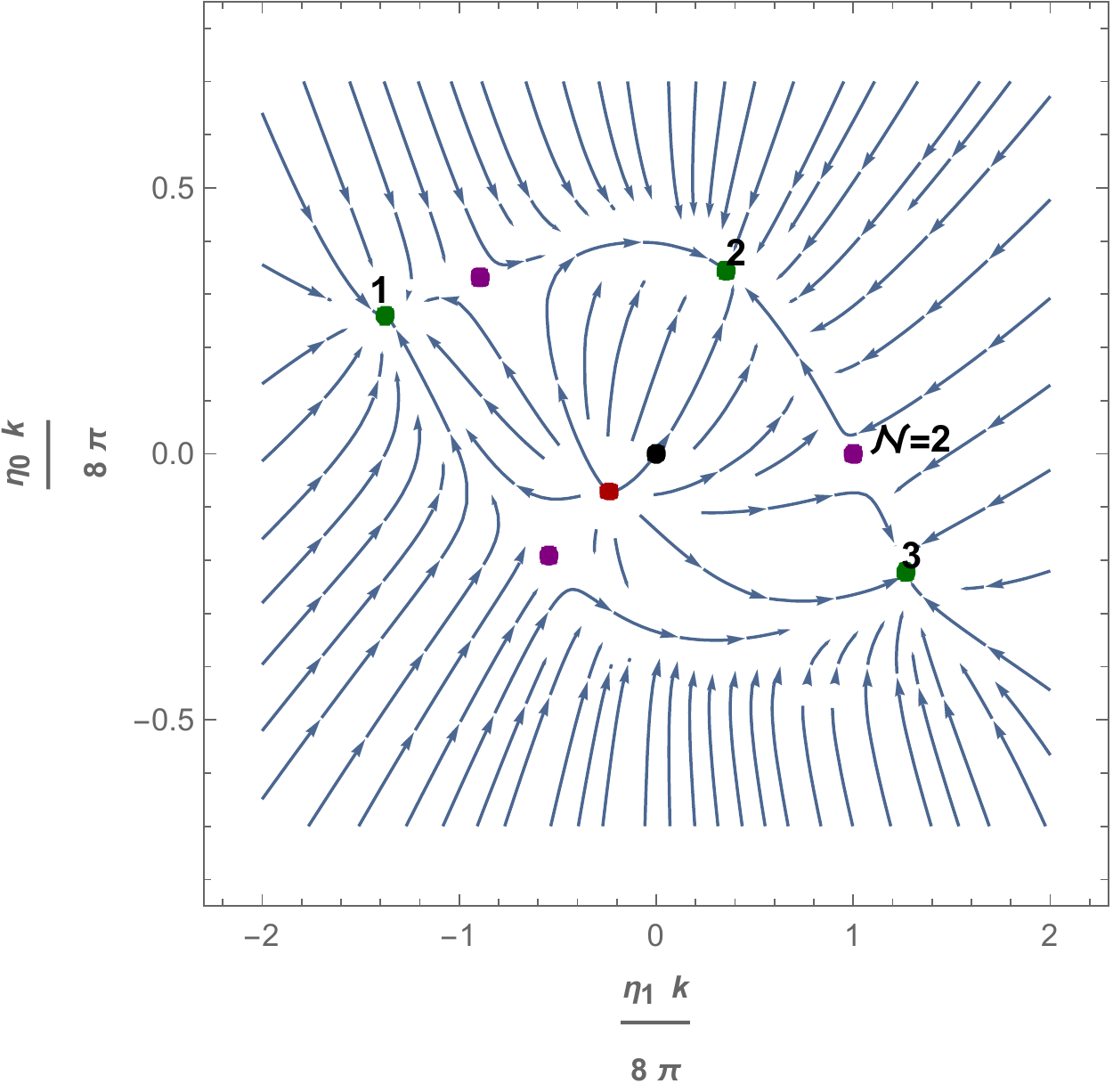}
    \caption{Renormalization group flows in $SU(4)_k$ theory with four fundamental flavors, at large $k$. Black marker denotes the origin, Red marker denotes a UV stable fixed point, green markers denote IR stable fixed points, and and purple markers denote saddle points. One of the saddle points possesses $\mathcal{N}\,=\,2$ supersymmetry.}
    \label{RGSU4}
\end{figure}
Above we have applied the same line of arguments as in section \eqref{Section3}, and again we were led to conclude that supersymmatry enhancement takes place. A natural check to perform at this point is to study the renormalization group flow near the second order phase transition point we have observed. At large CS level we can neglect the Yang-Mills term, and the role of gauge coupling is played by $g\,=\,\sqrt{\tfrac{8\pi}{k}}$ \cite{Avdeev:1991za, Avdeev:1992jt}. Therefore, at large $k$ the theory is weakly coupled (at least as far as the gauge sector is concerned), and so we can rely on perturbation theory. There are three classically marginal superpotential operators controlling the RG flow:
\begin{equation}
    \tfrac{1}{4}\eta_0\,(\text{Tr}\,\bar{Q}\,Q)^2\,+\,\tfrac{1}{4}\eta_0\,(\text{Tr}\,\bar{Q}\,T^A\,Q)^2\,+\,\tfrac{1}{4!}\,(\lambda\,\text{det}Q\,+\,\bar{\lambda}\,\text{det}\bar{Q}).
\end{equation}
Beta functions for these tree couplings take the form
\begin{align}
    \beta_{\eta_0}\,=\,\frac{3}{8192\pi^2}\,(175\eta_1^3\,+\,1440\eta_1^2\,\eta_0\,+\,2240\eta_1\,\eta_0^2\,+\,9216\eta_0^3\,+\,35\eta_1^2\,g^2\,+\,320\eta_0^2\,g^2\,-\nonumber\\
    -\,105\eta_1\,g^4\,-\,1280\eta_0\,g^4\,-\,105g^6\,+\,\tfrac{1}{3}\ |\lambda|^2\,(1920 g^2\,+\,33920\,\eta_0\,-\,1920\eta_1),\\
    \beta_{\eta_1}\,=\,\frac{1}{1024}\,(275\eta_1^3\,+\,1104\eta_1^2\eta_0\,+\,2752\eta_1\eta_0^2\,+\,240\eta_1\,\eta_0\,g^2\,+\,106\eta_1^2\,g^2\,-\,96\eta_0\,g^4\,-\nonumber\\
    -\,303\eta_1\,g^4\,-\,78\,g^6\,-\,|\lambda|^2\,(12080g^2\,+\,6144\eta_0\,-\,1280\eta_1)),\\
    \beta_{\lambda}\,=\,\frac{\lambda}{1024\pi^2}\,(768|\lambda|^2\,+\,71\eta_1^2\,-\,208\,\eta_0\,\eta_1\,+\,2752\eta_0^2\,+\,32\eta_0\,g^2\,+\,76\eta_1\,g^2\,+\,\frac{1225}{4}g^4),
\end{align}
where beta functions for $\lambda\,=\,0$ were known since \cite{Avdeev:1992jt}, and $\lambda$-dependent part is our contribution.

Let us first describe the situation when baryon deformation is turned off. RG trajectories for this case are depicted on fig. \eqref{RGSU4}. The fixed point relevant for the phase diagram under consideration is the green point number $2$ (as can be seen using the analysis applied in the previous sections, other IR fixed points describe transitions between the sets of vacua different from those appearing on the phase diagram for $SU(3)_k$ with three flavors, and preserved $U(1)_B$).

Turning on $\lambda\,\neq\,0$, we first notice that $\lambda$ itself is IR free, and the baryon operator is irrelevant in the IR. If we now deform an IR stable fixed point from fig. \eqref{RGSU4} by the baryon operator, we will flow to the same point we started with, or at most cam jump to some other fixed point from the same diagram, meaning that no supersymmetry enhancement can be observed.

The explanation of this discrepancy is that, as we have just stated, the baryon operator is IR free (we do not expect this fact to be modified also in the full theory with the Yang-Mills term included), so that we are dealing with an effective field theory, which is not reliable at large distances in the field space. In particular, we can not really trust the large mass analysis from which we conclude that there must be two vacua for large positive mass phase, and this piece of information was crucial for our arguments. It would be interesting to understand better the UV dynamics of this theory (or find its UV completion), permitting the reliable study of the phase diagram.

\section*{Acknowledgements} 
The work of VB is supported by the ERC Starting Grant 637844-HBQFTNCER and by the INFN. HK thanks University of Milano-Bicocca for hospitality and support during the completion of this work and Sara Pasquetti for making possible this visit.

\end{document}